\documentclass[journal,twoside,web]{ieeecolor2}

\makeatletter
\def\ps@titlepagestyle{%
  \def\@oddhead{}\def\@evenhead{}%
  \def\@oddfoot{}\def\@evenfoot{}%
}
\def\ps@headings{%
  \def\@oddhead{\hfill\thepage\hfill}%   just a bare page-number
  \def\@evenhead{\hfill\thepage\hfill}%
  \def\@oddfoot{}\def\@evenfoot{}%
}
\makeatother

\usepackage{generic}

\usepackage{amsmath,amssymb,amsfonts}
\usepackage{algorithmic}
\usepackage{graphicx}
\usepackage{textcomp}
\usepackage{orcidlink}
\usepackage{adjustbox}
\usepackage{booktabs}
\usepackage[style=numeric, maxnames=2, minnames=1, sorting=none]{biblatex}
\def\BibTeX{{\rm B\kern-.05em{\sc i\kern-.025em b}\kern-.08em
    T\kern-.1667em\lower.7ex\hbox{E}\kern-.125emX}}

{}
\begin{document}
\title{A Comparative Study of Conventional and Tripolar EEG for High-Performance Reach-to-Grasp BCI Systems}

%%%%%% Authors %%%%%%
\author{Ali Rabiee \orcidlink{0000-0001-6800-6247}, Sima Ghafoori \orcidlink{0000-0003-1866-5935}, Anna Cetera \orcidlink{0009-0003-4522-9220}, Maryam Norouzi \orcidlink{0000-0002-3253-9217}, Walter Besio \orcidlink{0000-0003-2752-5483}, \IEEEmembership{Senior Member, IEEE}, Reza Abiri \orcidlink{0000-0001-8975-8210}, \IEEEmembership{Member, IEEE}
\thanks{The research reported in this presentation was supported by the URI Foundation Grant on Medical Research. Additionally, the project was supported by the Rhode Island INBRE program from the National Institute of General Medical Sciences of the NIH under grant number P20GM103430, and by the NSF under award ID 2245558.}
\thanks{Ali Rabiee, Sima Ghafoori, Anna Cetera, Walter Besio, and Reza Abiri are with the Department of Electrical, Computer, and Biomedical Engineering, University of Rhode Island, Kingston, USA. Maryam Norouzi is with the Department of Mechanical Engineering, University of Rhode Island, Kingston, USA. (ali.rabiee@uri.edu; sima.ghafoori@uri.edu; annacetera@uri.edu; Maryam\_norouzi@uri.edu; besio@uri.edu; reza\_abiri@uri.edu).}
\thanks{First and Second authors contributed equally to this work.}
}

\maketitle

\begin{abstract}
\textit{Objective:} This study aims to enhance brain-computer interface (BCI) applications for individuals with motor impairments by comparing the effectiveness of noninvasive tripolar concentric ring electrode electroencephalography (tEEG) with conventional electroencephalography (EEG) technology. The goal is to determine which EEG technology is more effective in measuring and decoding different grasp-related neural signals. \textit{Methods:} The approach involves experimenting on ten healthy participants who performed two distinct reach-and-grasp movements: power grasp and precision grasp, with a no-movement condition serving as the baseline. Our research compares EEG and tEEG in decoding grasping movements, focusing on signal-to-noise ratio (SNR), spatial resolution, and wavelet time-frequency analysis. Additionally, our study involved extracting and analyzing statistical features from the wavelet coefficients, and both binary and multiclass classification methods were employed. Four machine learning algorithms—Random Forest (RF), Support Vector Machine (SVM), Extreme Gradient Boosting (XGBoost), and Linear Discriminant Analysis (LDA)—were used to evaluate the decoding accuracies. \textit{Results:} Our results indicated that compared with conventional EEG, tEEG yielded higher SNR, finer spatial resolution, and stronger wavelet power spectra. These advantages translated to superior decoding accuracies: tEEG reached around 90.00\% accuracy in binary classification and 75.97\% in multiclass tasks, versus 77.85\% and 61.27\% for conventional EEG. \textit{Conclusion:} tEEG provides richer and cleaner neural information than conventional EEG, enabling significantly better differentiation of grasp types. \textit{Significance:} These findings position tEEG as a promising alternative to conventional EEG for next-generation BCIs aimed at restoring or augmenting motor function in people with upper limb movement disabilities.
\end{abstract}


\section{Introduction}
\label{sec:introduction}
\IEEEPARstart{T}{he} development of brain-computer interfaces (BCIs) introduces a groundbreaking technology that facilitates direct interaction between the human brain and computer systems \cite{schalk2004bci2000, abiri2019comprehensive}. They offer substantial benefits to individuals with severe motor impairments, especially those resulting from spinal cord injuries, enhancing their quality of life through control over assistive devices and technologies \cite{rabiee2024streams, ghafoori2025novel}. A key and challenging aspect in BCI applications is decoding grasp-related neural activity. This isn't just about motor function; it integrates sensory perceptions, motor planning, and cognitive intentions \cite{castiello2008cortical}. The significance of accurately decoding these activities extends beyond scientific interest, holding the potential to revolutionize assistive technologies and facilitate natural interactions for BCI users.

The act of reaching and grasping is central to human motor functions and crucial for daily tasks. This involves various brain areas: the parietal regions, including the anterior intraparietal sulcus, the premotor regions like the ventral premotor cortex, and the primary motor cortex \cite{grafton2010cognitive, gallivan2013decoding}. While grasping plays a pivotal role in our interactions, decoding its underlying complex neuronal mechanisms remains a primary objective in neuroscience \cite{castiello2005neuroscience}. Achieving this can transform numerous domains: from enhancing rehabilitation medicine \cite{park2013rehabilitation, lambercy2009rehabilitation, thrasher2008rehabilitation} and advancing assistive technology \cite{grimm2016closed, rupp2014brain, millan2010combining, rabiee2024wavelet, ghafoori2024bispectrum} to designing intuitive brain-machine interfaces \cite{hazrati2010online, chang2019architecture}. Grasping requires precise muscle and joint coordination, adapting to the object's properties. Unraveling such intricate actions from brain signals is a challenging task. Many studies have explored the brain patterns linked to grasping actions in non-human primates \cite{donoghue1998neural, schaffelhofer2016object, michaels2018population} and humans using different acquisition modalities including invasive, such as electrocorticography (ECoG) \cite{jiang2020power} and noninvasive, such as functional magnetic resonance imaging (fMRI) \cite{fabbri2016disentangling, perini2020neural}. Additionally, numerous studies have demonstrated that noninvasive Electroencephalogram (EEG) signals can effectively decode reach-and-grasp actions.  In this context, machine learning algorithms are pivotal in decoding complex neural data. Deep learning models like Convolutional Neural Networks (CNNs) and Recurrent Neural Networks (RNNs) offer advanced analysis capabilities but usually require large datasets, which can be challenging to collect from human participants \cite{alzubaidi2023survey}. Therefore, many studies prefer traditional models such as SVMs, Decision Trees, and LDA. These traditional algorithms are more manageable with smaller datasets and still significantly improve the accuracy of movement decoding and prediction \cite{saeidi2021neural}. Iturrate et al \cite{iturrate2018human} utilized EEG to distinguish brain activation patterns during self-paced grasping tasks involving both precision and power grips, achieving 70\% accurate single-trial decoding by using shrinkage LDA (sLDA). Schwarz et al. \cite{schwarz2017decoding} demonstrated that it's possible to distinguish between three types of reach-and-grasp actions—palmar, pincer, and lateral grasp—using EEG neural correlates. The study achieved classification accuracies of up to 72.4\% between grasp types and 93.5\% for grasps versus no-movement conditions, supporting the potential use of this technique for more intuitive control of neuroprostheses. In another study by Xu et al \cite{xu2021decoding}, five distinct reach-and-grasp movements were decoded using movement-related cortical potentials (MRCPs) from noninvasive EEG signals. The movements included palmar, pinch, push, twist, and plug grasp. The study achieved a peak average accuracy of about 75\% for grasping vs. no-movement conditions. 
 
Despite the numerous studies utilizing EEG for decoding grasping movements, EEG often suffers from a low spatial resolution, complicating the precise localization of the source of neural activity due to the blurring effects primarily from different conductivities of the volume conductor \cite{nunez1994theoretical}. It is also vulnerable to physiological disturbances caused by muscle contractions, eye movements, and heart activity, as well as technical artifacts like power-line noises or fluctuations in electrode impedances \cite{jiang2019removal}. These limitations directly impact the decoding of grasp-related neural activity. Such a limitation emphasizes the need to explore advanced methods aimed at improving the spatial resolution of EEG data and increasing the signal-to-noise ratio. A promising solution is the use of tEEG, a novel technique involving an additional local reference electrode. The tripolar concentric ring electrode (TCRE) is roughly the same size as a conventional electrode. However, it consists of a central conductive disc with two electrically isolated rings surrounding it. The output signal from the TCRE is derived from a weighted sum of the outer ring subtracted from the central disc, which is then subtracted from sixteen times the difference between the inner ring and the central disc \cite{besio2006tri}. Compared to conventional disc electrode EEG, TCRE offers superior performance, with roughly 2.5 times higher spatial selectivity, 3.7 times higher signal-to-noise ratio, and approximately 12 times lower mutual information  \cite{koka2007improvement, aghaei2020tripolar}. There are other studies \cite{toole2019source, makeyev2017recent} that demonstrated tEEG could effectively locate high-frequency activity in epilepsy patients, predominantly in the seizure onset zone, suggesting its potential utility in identifying epileptic brain regions. Also in \cite{alzahrani2021comparison}, authors illustrated that tEEG significantly outperformed conventional disc electrodes in distinguishing individual finger movements, based on movement-related potentials (MRPs). Despite the potential benefits of tEEG, the use of tEEG in the realm of BCIs, particularly for decoding human grasping actions, remains largely unexplored.

\begin{figure}[ht!] %!t
\centering
\includegraphics[width=2.3 in]{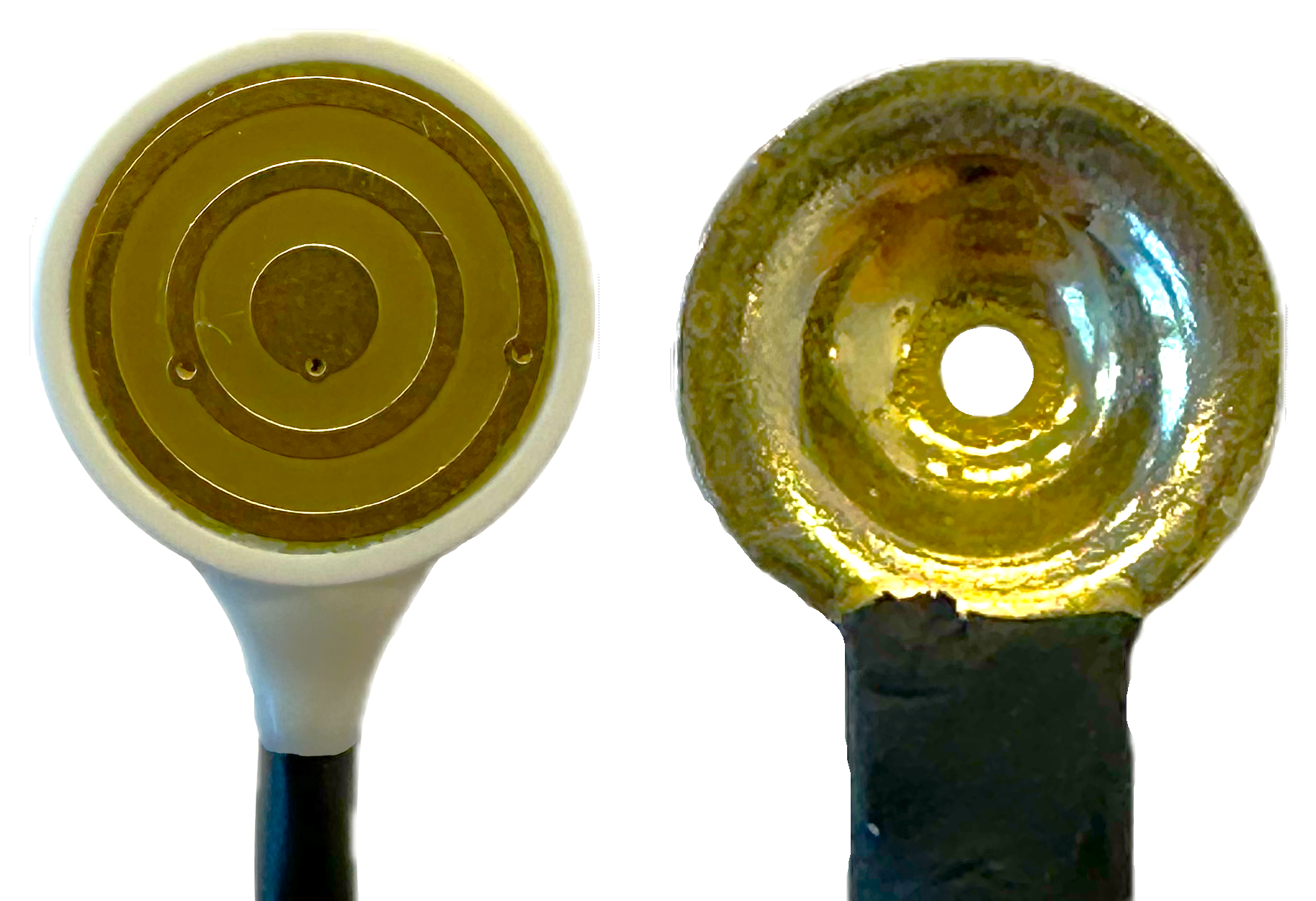}
\caption{(Left) tEEG electrode. (Right) Conventional EEG electrode.}
\label{Courant_2}
\end{figure}

In this study, we address critical challenges in decoding grasp-related neural activity and propose novel approaches to improve brain-computer interface (BCI) performance. The main contributions of this work are summarized as follows:

\begin{itemize}
    \item This research systematically compares tEEG and conventional EEG technologies for grasp-related tasks, demonstrating significant improvements not only in SNR and spatial resolution but also in decoding accuracy with tEEG, particularly in both binary and multiclass classification scenarios.
    
    \item The use of wavelet-based techniques alongside tEEG electrodes improves the interpretation of neural activity and decoding accuracy, offering deeper insights into the dynamic nature of grasp-related brain signals.
    
    \item Despite utilizing only four electrodes, traditional machine learning algorithms, and a cost-effective EEG headset, the study achieves higher classification accuracies compared to existing methods, demonstrating the feasibility of achieving high performance with minimal hardware and computational resources.
    
    \item The features extracted using wavelet-based techniques and tEEG signals are demonstrated to be rich and robust, enabling consistent performance across various machine learning algorithms.
\end{itemize}

\begin{figure*}[ht!] %!t
\centering
\includegraphics[width=6 in]{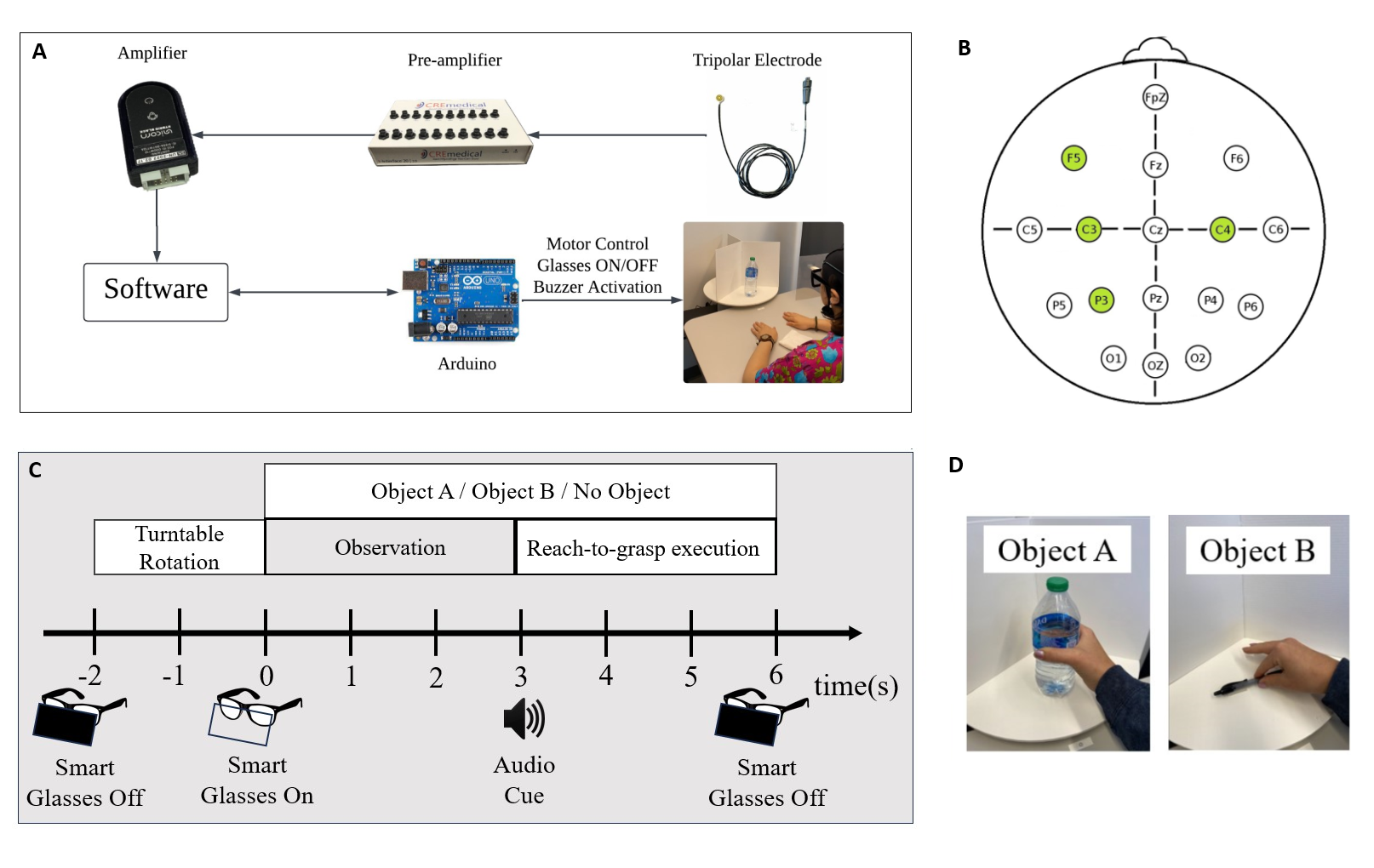}
\caption{Experimental setup and paradigm for reach-and-grasp tasks. (A) Platform components and connections, (B) an illustration of a 16-channel EEG sensor layout with the 4 electrodes(in green) which are used in the study to decode grasp movements, and (C) the experimental paradigm based on the audio cue. (D) Objects A and B are randomly selected from a bottle and a pen that require power and precision grasps, respectively.}
\label{Courant_2}
\end{figure*}

%%%% Methods %%%%
\section{Materials and Methods}
A total of ten healthy volunteers, aged between 22 and 35 years consisting of five males and five females, participated in this study. To determine the requisite sample size for our study, a preliminary power analysis was conducted using G*Power \cite{kang2021sample}, based on our pilot data. This analysis aimed to identify the number of participants needed to detect a statistically significant difference in decoding accuracies between two conditions, and the results indicated that a total of 10 participants would be sufficient for this purpose.  The study was non-invasive and received approval from the University of Rhode Island Institutional Review Board (IRB, \#1944644-4 on January 4, 2024). All participants had no known prior medical conditions and possessed normal or corrected-to-normal vision. They were thoroughly briefed about the study's objectives. Before initiating any recordings, each participant provided informed consent, ensuring they understood the research's purpose and procedures. 

\subsection{Experimental Setup and Paradigm}

To study the neural activity associated with natural vision-driven grasping tasks, we established a novel and comprehensive experimental setup and study protocol. As illustrated in Figure 2A, the setup includes tripolar electrodes, a t-interface pre-amplifier, a g.tec Unicorn Hybrid Black amplifier, an Arduino, and custom software developed to manage the experiment and integrate various hardware components. A key feature of the setup is a 3D-engineered, motorized turntable divided into three segments, which display two objects (a bottle and a pen) or an empty space for a no-object scenario. Participants were seated comfortably with their hands facing down, positioned 30 cm from the object’s center. Following a three-second observation period accompanied by a buzzing sound signal, participants initiated the grasping phase. The turntable then rotated randomly to the next condition—object A, object B, or no-object—while their eyeglasses turned opaque to prepare for the subsequent phase (Figure 2C, 2D).

A notable advancement in our experimental design is the incorporation of these innovative smart eyeglasses, equipped with a special film that alternates between clear and opaque states developed in our previous study \cite{cetera2024classification}. This feature is specifically designed to reduce potential anticipatory bias by obscuring the participants' view of the objects until the experimental condition is fully set. This addition addresses a known challenge in studies of neural activity, as prior findings, such as those discussing the Fehrer-Raab effect \cite{fehrer1962reaction, breitmeyer2006visual, schurger2010reproducibility}, suggest that neural activities can begin before a target becomes fully visible, influenced by prior context or subtle visual cues. In our setup, neural activities might initiate even before the turntable completes its rotation, as participants could anticipate the upcoming condition. By leveraging visual masking through the eyeglasses, our approach minimizes this anticipatory activity, ensuring that the recorded neural signals more accurately reflect task-relevant processes associated with the actual execution phase.

\subsection{Data Acquisition}

EEG data acquisition was realized using the Unicorn Hybrid Black Bluetooth amplifier, capturing data at a rate of 250 samples per second per channel. Prior to initiating each session, skin-to-electrode impedance was assured to be below 10k$\Omega$ by impedance meter to optimize data integrity. While Alzahrani's work \cite{alzahrani2021comparison} demonstrated that only a single TCRE electrode at the C3 motor area location was sufficient for classifying imaginary finger movements, our study extends this approach by also considering additional brain regions such as parietal, central, and frontal areas, which play distinct and vital roles in the various stages of grasping tasks, from planning to execution \cite{vingerhoets2014contribution}. To this end, we have placed four TCREs at the P3, C3, C4, and F5 locations based on the 10-20 system as illustrated in Figure 2B, concurrently capturing tEEG and conventional EEG data \cite{makeyev2013emulating}. Reference and ground electrodes were fixed at the left and right mastoids, respectively. Each participant underwent 50 trials for each condition, allowing us to gather a detailed and extensive dataset for further analysis.

\subsection{Preprocessing}

Data analysis was executed utilizing custom Python scripts, leveraging functions from the mne package \cite{gramfort2013meg}. Initially, the data were bandpass filtered using a Finite Impulse Response (FIR) filter of order 20, spanning a frequency range of 1Hz—60Hz \cite{widmann2015digital, widmann2012filter}.  Post the spectral filtering, wavelet denoising, an advanced signal processing technique, was implemented to further refine the quality of the EEG recordings. As shown by the authors in \cite{carmona1994wavelet}, wavelet denoising can be beneficial in EEG signals by removing noise while preserving salient features associated with underlying neuronal activities. For our denoising procedure, we utilized the Daubechies wavelet, specifically the db8 waveform, known for its localization properties in both the time and frequency domains \cite{daubechies1992ten}. The db8 wavelet has been identified as the most effective wavelet for noise reduction in EEG signals when analyzing data from healthy subjects \cite{mamun2013effectiveness}. This particular wavelet is distinguished by its ability to eliminate extraneous noise, enhancing the clarity and reliability of EEG readings\cite{asaduzzaman2010study}.

With the denoised signals in place, the subtraction of the mean reference and a detrending process was performed, aiming to counteract the introduction of spurious trends in the spectral data \cite{de2018robust}. To ensure a consistent scale across data, the signals were subjected to Z-score normalization. This technique is essential for ensuring that EEG data is uniform across different measurements, which significantly contributes to improving the accuracy and effectiveness of both traditional machine learning and advanced deep learning models \cite{shoeibi2021automatic}.

For a focused examination of movement-related activity, the data were meticulously segmented to capture the temporal window spanning 1 second before to 1 second after movement onset. Culminating the preprocessing pipeline, an Independent Component Analysis (ICA) was administered \cite{hyvarinen2000independent, bell1995information}. This technique identified and eliminated components tainted by physiological or non-physiological noise extraneous to EEG signals, for subsequent data analysis.

\subsection{Comparative Analysis}

The foundational aspect of this study revolves around a comprehensive comparison between conventional EEG and tEEG technologies. Before delving into classification-based analyses, it is crucial to discern the inherent differences and potential advantages one technology may offer over the other. This section presents the methodologies employed to undertake such comparative analyses, focusing on three pivotal techniques: SNR, functional connectivity, and Time-Frequency analysis using wavelets. \\
\subsubsection*{Signal-to-Noise Ratio (SNR)}

In evaluating the efficacy of our preprocessing steps, we analyzed the SNR for both EEG and tEEG configurations across multiple channels while participants performed the grasping tasks. This analysis aimed to verify if the tEEG setup indeed offers enhanced signal quality compared to the standard EEG. For calculating SNR, we compared the wavelet-denoised signal to the noise, which is determined as the difference between the original signal and its denoised version. The SNR is computed using the formula:
\[ \text{SNR} = 20 \times \log_{10}\left(\frac{\text{RMS of the Signal}}{\text{RMS of the Noise}}\right) \]
Here, RMS represents the Root Mean Square. In the context of a continuous waveform, the RMS is the square root of the mean power of the signal. For a discrete sequence, it refers to the square root of the sum of the squares over the total data points. Both EEG and tEEG data underwent SNR computation across all channels. It is noteworthy to mention that the presented results represent grand averages accumulated from all subjects, ensuring a comprehensive overview. The data used for this analysis was specifically segmented around the peri-movement time frame. \\

\subsubsection*{Frequency-Domain Functional Connectivity Analysis}

Functional connectivity among channels was assessed using the coherence method, which quantifies the degree of synchrony between pairs of signals in the frequency domain. Specifically, coherence elucidates the linear time-invariant relationship between two signals, offering insights into how distinct channels may be interacting or how different brain regions may be synchronized.

The data, obtained from the tEEG setup and organized by channels and time points, underwent a coherence analysis for each possible pair of channels. Given that coherence is computed across various frequencies, an average value was derived for each channel pair to represent their overall synchrony. This resulted in a coherence matrix, which is symmetric by design, capturing the relationships among all channels. \\

\subsubsection*{Time-Frequency Analysis}
 
Building upon the foundational analyses of SNR and functional connectivity, we extend our methodological framework to encompass time-frequency analysis by using wavelets. While features derived from the time domain, provide valuable insights into neural dynamics, they offer a limited perspective on the intricate oscillatory nature of brain signals \cite{cohen2014analyzing}. To capture a more comprehensive feature map, it is imperative to consider both the time and frequency characteristics of the EEG and tEEG data.

We employed time-frequency analysis using wavelets, which allows for the decomposition of EEG and tEEG signals into time and frequency domains simultaneously. This approach is particularly adept at identifying non-stationary signals, which are common in EEG data associated with motor tasks. For our analysis, we utilized the Morlet wavelet, a commonly used wavelet in neuroscience due to its biological relevance and ability to provide a good balance between time and frequency localization. The continuous wavelet transform (CWT) was applied to the preprocessed signals from both EEG and tEEG data to obtain the time-frequency representation of the underlying neural activity. 

In our computational approach, we first determined the power of the wavelet coefficients by calculating the absolute value to find the energy content at each frequency and time point. Subsequently, we averaged these power values across all trials to yield a mean spectral density for each channel. The resulting average power spectra were visualized as heatmaps, which allowed for a direct comparison between the EEG and tEEG data. These visual representations were scaled to facilitate a clear comparison, with the color intensity indicating the power at each time-frequency point.

The average power spectra, visualized as heatmaps, serve not only as a tool for feature extraction but also as a means for a detailed comparison between EEG and tEEG. The heatmaps facilitate a side-by-side evaluation of how EEG and tEEG represent neural dynamics across different frequency bands and time intervals. This comparative analysis allows us to discern which modality provides a richer and more detailed depiction of brain activity, thereby offering more informative features for the classification of motor tasks.

\subsection{Feature Extraction and Classification}

The primary objective during the feature extraction phase was to capture relevant information from the EEG and tEEG data. The time window [-1, 1] s was defined as the time region of interest (tROI), with 0 s corresponding to the movement onset. Wavelets, specifically the complex Morlet wavelet (cmor), were employed to decompose these segments into various frequency bands: delta, theta, alpha, beta, and gamma. For each of the identified frequency bands in every channel, four statistical features were extracted from the magnitude of the wavelet coefficients. These features are: mean of the magnitude, variance of the magnitude, skewness of the magnitude, and kurtosis of the magnitude.

Given that there were 4 channels, each channel containing data from 5 frequency bands, and each band producing 4 features, the total feature count amounted to 80 for every data instance. With the features in hand, the subsequent step was classification. We employed four distinct machine learning algorithms: Support Vector Machine (SVM), Random Forest (RF), Extreme Gradient Boosting (XGBoost), and Linear Discriminant Analysis (LDA). These techniques were used for both multiclass and binary classification scenarios. To assess the effectiveness of these models, we implemented a 5-fold cross-validation method. This approach helped us gain a reliable measure of the models' ability to generalize, as it calculated the average performance indicators across all the validation folds. For all comparisons and analyses in this study, paired t-tests were employed to statistically validate the results.

\section{Results}

\subsection{Comparative Analysis}

The comparative analysis of SNRs between the tEEG and conventional EEG technologies revealed a consistent pattern favoring the tEEG method across various channels. As delineated in Figure 3, the grand average SNR values for all subjects besides the p-values were assessed at channels P3, C3, C4, and F5. Notably, the tEEG (green boxes) significantly exhibited higher SNRs across all channels when compared to the conventional EEG (blue boxes). Specifically, the median SNR for the tEEG ranged from 56.24 dB at channel C3 to 59.56 dB at channel P3, which was the highest observed across all channels and technologies. In comparison, the conventional EEG technology presented lower median SNRs, with the lowest at channel C4 (53.92 dB) and the highest at channel P3 (54.57 dB). In addition to the improved median SNRs, the tEEG also demonstrated lower variability in SNR measurements among subjects, as evidenced by the narrower interquartile ranges in the green boxes. This reduced variance suggests that the tEEG provides more consistent and reliable SNR values, which is a critical aspect of EEG signal quality.

\begin{figure}[ht!] %!t
\centering
\includegraphics[width=3.5 in]{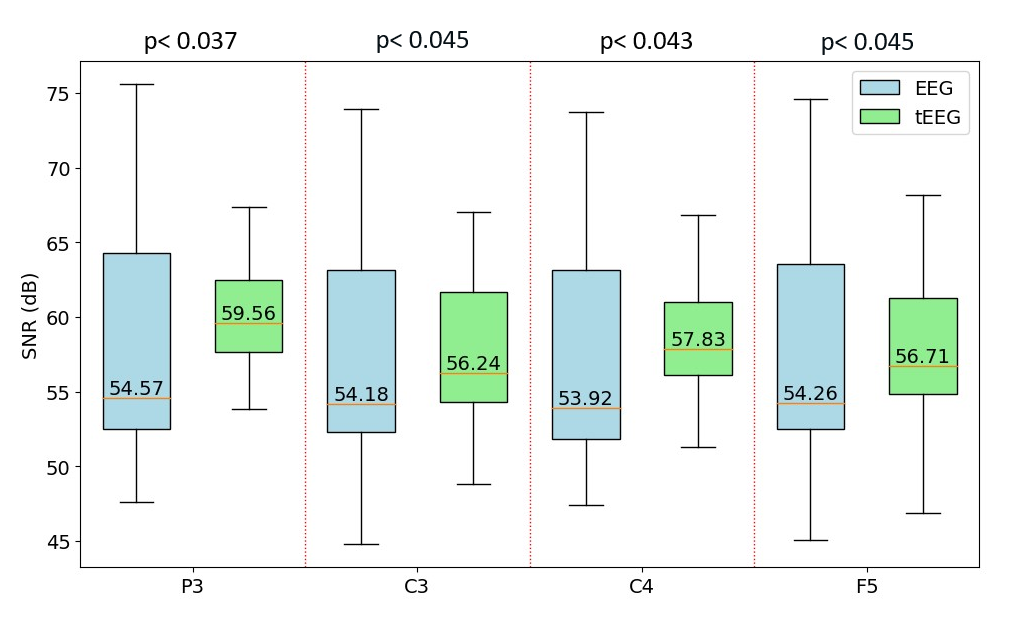}
\caption{Box plot representation of grand average SNR across P3, C3, C4, and F5 channels, comparing tEEG (green) to EEG (blue).}
\label{Courant_2}
\end{figure}
\begin{figure}[ht!] %!t
\centering
\includegraphics[width=3.5 in]{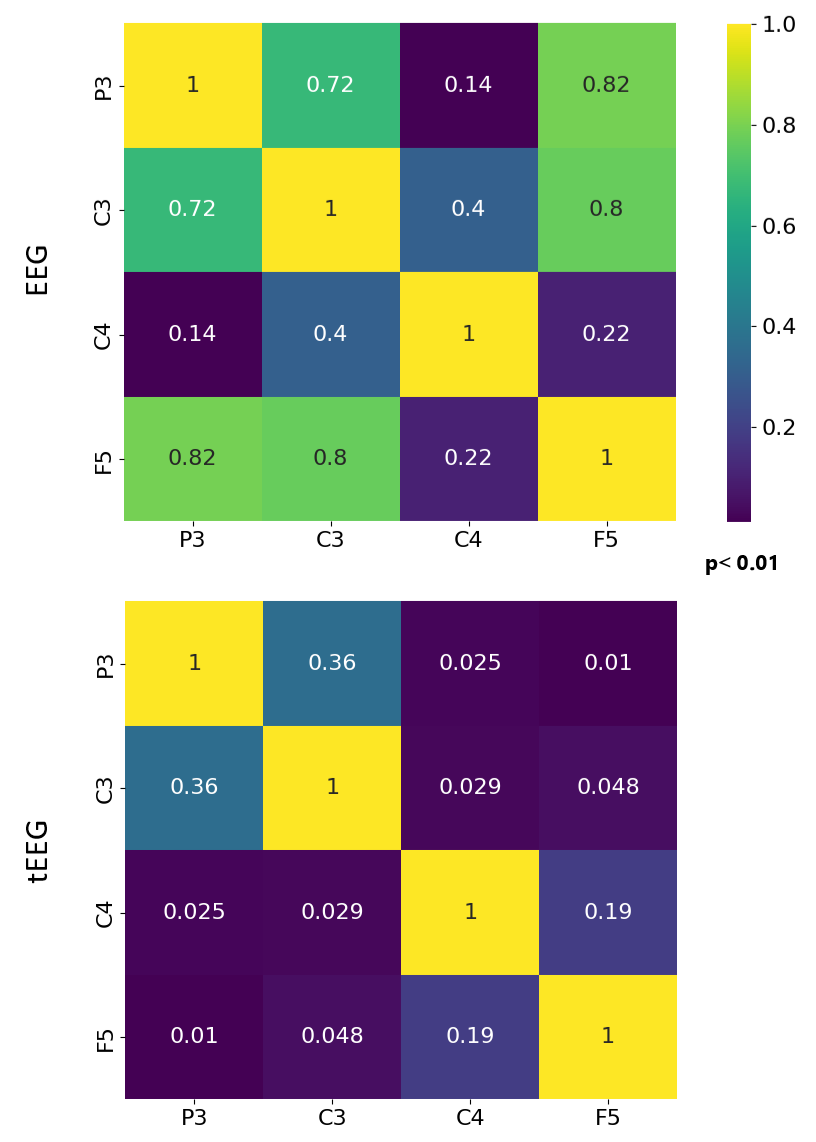}
\caption{Heatmaps of grand average functional connectivity across EEG channels for EEG (top) and tEEG (bottom).}
\label{Courant_2}
\end{figure}

\begin{figure*}[ht!] %!t
\centering
\includegraphics[width=6 in]{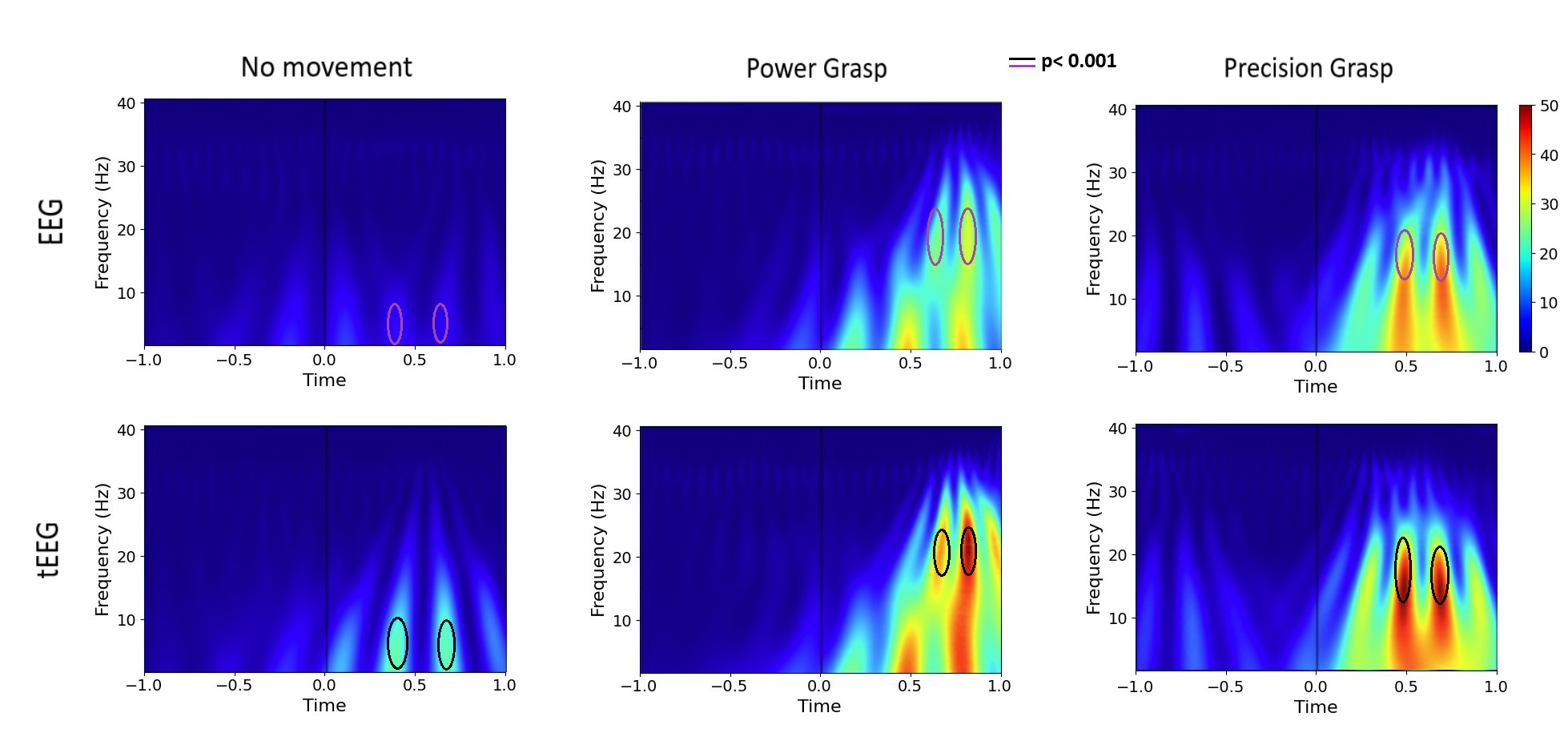}
\caption{ Wavelet time-frequency maps comparing EEG (left column) to tEEG (right column) for Power Grasp (top row), Precision Grasp (middle row), and No Movement (bottom row) conditions.}
\label{Courant_2}
\end{figure*}
\begin{table*}[h]
\centering
\caption{Binary Classification Accuracies for No-movement vs. Power Grasp Tasks Using Various Models, Comparing EEG and tEEG Accuracies Across Different Subjects.}
\label{tab:results}
\begin{adjustbox}{width=\textwidth,center}
\begin{tabular}{lcccccccc}
\toprule
Subjects & \multicolumn{2}{c}{Random For. (\%)} & \multicolumn{2}{c}{SVM (\%)} & \multicolumn{2}{c}{XGBoost (\%)} & \multicolumn{2}{c}{LDA (\%)} \\
\cmidrule(lr){2-3} \cmidrule(lr){4-5} \cmidrule(lr){6-7} \cmidrule(lr){8-9}
 & EEG & tEEG & EEG & tEEG & EEG & tEEG & EEG & tEEG \\
\midrule
S1 & 70.00& 90.00& 75.00& 75.56 & 73.33 & 88.89 & 51.67 & 78.89 \\
S2 & 82.82 & 87.42 & 76.54 & 85.42 & 84.36 & 84.32 & 56.28 & 72.89 \\
S3 & 71.76 & 92.36 & 74.05 & 83.3 & 76.34 & 90.17 & 50.26 & 84.10\\
S4 & 95.00& 96.11 & 89.17 & 85.56 & 90.83 & 97.22 & 68.33 & 76.67 \\
S5 & 69.58 & 90.56 & 69.17 & 68.06 & 67.08 & 86.11 & 60.42 & 69.44 \\
S6 & 80.00& 91.33 & 73.00& 78.67 & 76.00& 88.67 & 43.00& 78.67 \\
S7 & 94.38 & 98.24 & 93.12 & 96.67 & 92.50& 98.33 & 84.38 & 89.58 \\
S8 & 73.75 & 89.58 & 70.00& 70.00& 71.25 & 90.00& 66.88 & 73.75 \\
S9 & 73.12 & 91.67 & 67.50& 72.92 & 65.62 & 90.00& 59.38 & 72.08 \\
S10 & 68.12 & 87.08 & 66.88 & 75.83 & 68.12 & 82.50& 60.00& 67.50\\
\midrule
Mean &  \(77.85 \pm 9.47\)& \(91.43 \pm 3.32\)& \(75.44 \pm 8.46\)& \(79.20 \pm 8.21\)& \(76.54 \pm 9.16\)& \(89.62 \pm 4.76\)& \(60.06 \pm 10.86\)& \(76.36 \pm 6.40\)\\
\bottomrule
\end{tabular}
\end{adjustbox}
\end{table*}
\begin{table*}[h]
\centering
\caption{Binary Classification Accuracies for No-movement vs. Precision Grasp Tasks Using Various Models, Comparing EEG and tEEG Accuracies Across Different Subjects.}
\label{tab:results}
\begin{adjustbox}{width=\textwidth, center}
\begin{tabular}{lcccccccc}
\toprule
Subjects & \multicolumn{2}{c}{Random For. (\%)} & \multicolumn{2}{c}{SVM (\%)} & \multicolumn{2}{c}{XGBoost (\%)} & \multicolumn{2}{c}{LDA (\%)} \\
\cmidrule(lr){2-3} \cmidrule(lr){4-5} \cmidrule(lr){6-7} \cmidrule(lr){8-9}
 & EEG & tEEG & EEG & tEEG & EEG & tEEG & EEG & tEEG \\
\midrule
S1 & 76.67 & 90.00 & 78.33 & 85.56 & 78.33 & 90.00 & 51.67 & 71.11 \\
S2 & 78.08 & 88.53 & 62.56 & 82.21 & 67.18 & 84.37 & 53.21 & 76.11 \\
S3 & 75.03 & 93.22 & 62.35 & 82.56 & 66.99 & 93.19 & 48.95 & 71.20 \\
S4 & 93.33 & 92.78 & 88.33 & 82.22 & 91.67 & 92.22 & 64.17 & 77.78 \\
S5 & 70.42 & 88.89 & 72.08 & 76.94 & 69.17 & 84.72 & 63.33 & 69.72 \\
S6 & 80.00 & 88.67 & 67.00 & 87.33 & 75.00 & 90.67 & 62.00 & 76.00 \\
S7 & 93.75 & 99.17 & 92.50 & 94.58 & 93.12 & 97.92 & 80.62 & 92.92 \\
S8 & 68.12 & 89.58 & 70.62 & 73.33 & 64.38 & 88.75 & 54.37 & 67.50 \\
S9 & 70.62 & 89.17 & 68.12 & 77.08 & 67.50 & 88.75 & 65.00 & 70.42 \\
S10 & 71.25 & 87.08 & 69.38 & 77.92 & 67.50 & 86.67 & 54.37 & 77.08 \\
\midrule
Mean & 77.73 \(\pm\) 8.67& 90.71 \(\pm\) 3.34& 73.13 \(\pm\) 9.72& 81.97 \(\pm\) 5.84& 74.08 \(\pm\) 9.97& 89.73 \(\pm\) 3.89& 59.77 \(\pm\) 8.85& 74.98 \(\pm\) 6.86\\
\bottomrule
\end{tabular}
\end{adjustbox}
\end{table*}
\begin{table*}[h]
\centering
\caption{Binary Classification Accuracies for Power Grasp vs. Precision Grasp Tasks Using Various Models, Comparing EEG and tEEG Accuracies Across Different Subjects.}
\label{tab:results}
\begin{adjustbox}{width=\textwidth, center}
\begin{tabular}{lcccccccc}
\toprule
Subjects & \multicolumn{2}{c}{Random For. (\%)} & \multicolumn{2}{c}{SVM (\%)} & \multicolumn{2}{c}{XGBoost (\%)} & \multicolumn{2}{c}{LDA (\%)} \\
\cmidrule(lr){2-3} \cmidrule(lr){4-5} \cmidrule(lr){6-7} \cmidrule(lr){8-9}
 & EEG & tEEG & EEG & tEEG & EEG & tEEG & EEG & tEEG \\
\midrule
S1 & 65.00 & 82.22 & 73.33 & 76.67 & 63.33 & 78.89 & 40.00 & 67.78 \\
S2 & 64.10 & 87.47 & 67.18 & 71.79 & 61.03 & 84.32 & 48.21 & 72.74 \\
S3 & 73.86 & 89.37 & 78.30 & 79.60 & 71.57 & 87.92 & 60.26 & 76.47 \\
S4 & 90.83 & 95.56 & 90.00 & 80.56 & 87.50 & 94.44 & 64.17 & 66.11 \\
S5 & 72.92 & 86.94 & 72.50 & 65.56 & 71.25 & 85.83 & 62.50 & 66.39 \\
S6 & 78.00 & 89.33 & 69.00 & 76.00 & 75.00 & 88.67 & 61.00 & 69.33 \\
S7 & 91.88 & 97.92 & 88.75 & 85.83 & 90.00 & 97.08 & 80.00 & 77.08 \\
S8 & 69.38 & 85.83 & 73.12 & 64.17 & 75.00 & 85.83 & 53.75 & 71.25 \\
S9 & 69.38 & 82.92 & 70.62 & 72.92 & 66.25 & 82.08 & 54.37 & 64.17 \\
S10 & 70.62 & 81.25 & 72.50 & 69.17 & 65.00 & 80.83 & 47.50 & 64.58 \\
\midrule
Mean & 74.60 \(\pm\) 9.22& 87.88 \(\pm\) 5.20& 75.53 \(\pm\) 7.47& 74.23 \(\pm\) 6.5& 72.59 \(\pm\) 9.26& 86.59 \(\pm\) 4.45& 57.18 \(\pm\) 10.54& 69.59 \(\pm\) 4.43\\
\bottomrule
\end{tabular}
\end{adjustbox}
\end{table*}
\begin{table*}[h]
\centering
\caption{Multiclass Classification Accuracies Across Different Subjects for Three Types of Tasks, Comparing EEG and tEEG Performance for Various Models.}
\label{tab:results}
\begin{adjustbox}{width=\textwidth, center}
\begin{tabular}{lcccccccc}
\toprule
Subjects & \multicolumn{2}{c}{Random For. (\%)} & \multicolumn{2}{c}{SVM (\%)} & \multicolumn{2}{c}{XGBoost (\%)} & \multicolumn{2}{c}{LDA (\%)} \\
\cmidrule(lr){2-3} \cmidrule(lr){4-5} \cmidrule(lr){6-7} \cmidrule(lr){8-9}
 & EEG & tEEG & EEG & tEEG & EEG & tEEG & EEG & tEEG \\
\midrule
S1 & 38.67 & 63.70 & 57.33 & 51.85 & 41.33 & 79.26 & 20.00 & 72.59 \\
S2 & 36.25 & 70.91 & 56.25 & 58.28 & 47.50 & 76.31 & 33.75 & 63.94 \\
S3 & 47.27 & 76.78 & 60.91 & 68.12 & 55.45 & 76.29 & 37.27 & 71.77 \\
S4 & 63.33 & 75.19 & 72.00 & 67.04 & 68.00 & 87.04 & 43.33 & 75.56 \\
S5 & 42.00 & 68.89 & 61.00 & 57.22 & 46.33 & 65.00 & 46.33 & 50.56 \\
S6 & 53.60 & 72.44 & 59.20 & 63.11 & 58.40 & 79.11 & 35.20 & 61.33 \\
S7 & 63.00 & 84.72 & 68.50 & 73.89 & 66.50 & 82.22 & 69.00 & 70.83 \\
S8 & 42.50 & 71.67 & 60.00 & 49.17 & 48.00 & 73.89 & 39.50 & 60.56 \\
S9 & 40.00 & 70.00 & 60.00 & 58.06 & 43.50 & 73.61 & 46.00 & 53.61 \\
S10 & 44.00 & 70.83 & 57.50 & 54.44 & 46.00 & 66.94 & 38.00 & 53.89 \\
\midrule
Mean & 47.06 \(\pm\) 9.24& 72.51 \(\pm\) 5.27& 61.27 \(\pm\) 4.80& 60.12 \(\pm\) 7.40& 52.10 \(\pm\) 9.00& 75.97 \(\pm\) 6.28& 40.84 \(\pm\) 11.81& 63.46 \(\pm\) 8.49\\
\bottomrule
\end{tabular}
\end{adjustbox}
\end{table*}

Figure 4 presents the grand average functional connectivity between EEG channels, quantified across all subjects for both EEG and tEEG technologies. The heatmaps visually compare the connectivity strengths, with conventional EEG displayed on the top and tEEG on the bottom. The conventional EEG heatmap shows considerable functional connectivity, with several channel pairs exhibiting strong correlations, such as P3-F5 (0.82) and C3-F5 (0.8), suggesting significant shared signal components across these channels. This is indicative of a lower spatial resolution, where the activity from different brain regions may be detected simultaneously by multiple channels. Conversely, the tEEG heatmap illustrates substantially lower connectivity values for all channel pairs when compared to conventional EEG. This is particularly evident in the markedly reduced correlations for channel pairs P3-F5 (0.01) and C3-F5 (0.048). 

Figure 5 showcases the wavelet time-frequency representations for conventional EEG and tEEG, analyzing the neural dynamics from one second before to one second after the onset of movement, which is marked by time zero on the x-axis. During the Power Grasp and Precision Grasp conditions, the time-frequency maps exhibit a high increase in activity, particularly noticeable around 500ms post-movement onset. The tEEG maps display a more significant increase in activity across a range of frequencies when compared to conventional EEG. This increased activity, especially evident in the beta frequency band, is indicative of the sensorimotor processing associated with the execution of the grasping tasks. In the No Movement condition, conventional EEG exhibits minimal activity, aligning with expectations of neural quiescence. However, tEEG reveals subtle low-frequency activities even in this resting state, indicating its heightened sensitivity to neural oscillations that conventional EEG may not capture.

\subsection{Feature Extraction and Classification}

Table 1 presents the binary classification accuracies for distinguishing between No-movement and Power Grasp tasks across different subjects using various machine learning models with accuracies reported for both conventional EEG and tEEG. Across all subjects, tEEG consistently outperformed conventional EEG in classification accuracy. The mean accuracy improvements with tEEG ranged from 3.76\% in SVM to a notable 16.30\% in LDA, indicating a substantial enhancement in the machine learning models' ability to utilize the data captured by tEEG. The Random Forest classifier achieved the highest mean accuracy for tEEG at 91.53\%. Similarly, LDA showed the least improvement, yet still produced a mean accuracy increase of 16.30\% with tEEG. Individual subject analysis revealed that Subject 7 (S7) exhibited the highest classification accuracy with tEEG at 98.33\% using the XGBoost model, while the lowest was for Subject 10 (S10) at 67.5\% using LDA. 

Table 2 details the binary classification accuracies for No-movement versus Precision Grasp tasks, utilizing a range of machine learning models and comparing the performance between conventional EEG and tEEG. The data illustrate a consistent trend where tEEG provides a marked improvement in classification accuracy over conventional EEG. The mean accuracies are significantly higher with tEEG, exhibiting an increase of 12.98\% for Random Forest, 8.84\% for SVM, 15.65\% for XGBoost, and 15.21\% for LDA. The XGBoost classifier shows the most substantial average increase in accuracy when using tEEG. Notably, every subject experienced an accuracy improvement with tEEG, affirming its universal enhancement of classification performance. Subject 7's results stand out with an exceptional 99.17\% accuracy using the Random Forest classifier in conjunction with tEEG, while the lowest accuracy with this technology is seen in Subject 8, with 67.50\% using LDA. This spread suggests that while tEEG generally improves classification outcomes, the extent of its impact can vary among individuals.

Table 3 summarizes the binary classification accuracies for discriminating between Power Grasp and Precision Grasp tasks, utilizing various machine learning models for both conventional EEG and tEEG. A key observation from the table is that while tEEG demonstrates an overall improvement in classification accuracy for most models, the SVM model with conventional EEG data exhibits a slightly higher average accuracy (75.53\%) compared to tEEG (74.23\%). However, this trend does not hold across other models. The XGBoost classifier, in particular, shows a marked increase in mean accuracy from 72.59\% with EEG to 86.59\% with tEEG. Subject 7's classification with tEEG using Random Forest underscores this potential. For Random Forest and LDA models, tEEG also outperforms conventional EEG, with mean accuracies higher by 13.28\% and 12.41\%, respectively. 
\begin{figure}[ht!] %!t
\centering
\includegraphics[width=3.5 in]{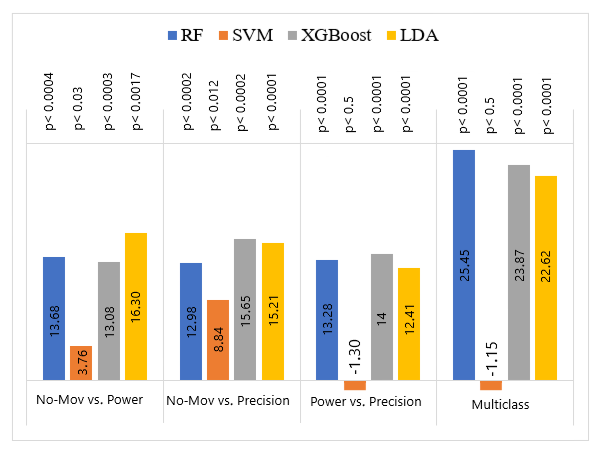}
\caption{Percentage improvement in classification accuracy using tEEG over conventional EEG for binary and multiclass tasks, across machine learning models.}
\label{Courant_2}
\end{figure}

Table 4 summarizes the classification accuracies for a multiclass scenario involving No-movement versus different grasp tasks, contrasting the performance between EEG and tEEG across various subjects and machine learning models. It is notable that almost all classification results exceed the chance level of 33.33\%, indicating that both EEG and tEEG provide signals containing sufficient information to distinguish between the classes above a level of random guessing. This is an important baseline as it establishes that the neural signals being classified carry task-relevant information. tEEG consistently shows an advantage over conventional EEG across all subjects and models, with the mean accuracy for tEEG being significantly higher. The Random Forest classifier in particular reveals a substantial mean accuracy increase from 47.06\% with conventional EEG to 72.51\% with tEEG. The performance of the SVM model exhibits robustness across both EEG and tEEG modalities, with nearly identical average results, although EEG displays a marginally higher accuracy. XGBoost and LDA models further reinforce this trend, showing substantial improvements when using tEEG data, with mean accuracies of 75.97\% and 63.46\% respectively. 

Figure 6 illustrates the improvement in classification accuracy achieved by employing tEEG over conventional EEG for various binary and multiclass tasks. The corresponding p-values obtained from statistical analysis using a paired t-test are also presented, indicating the statistical significance of the observed performance differences.  Each bar represents the percentage increase in accuracy across different machine learning models. The data suggest that while SVM is relatively robust, with smaller improvements seen with tEEG in some scenarios, the other models exhibit significantly accuracy enhancements when employing tEEG. Specifically, the improvement is more pronounced for RF, XGBoost, and LDA models. 

\section{Discussion}

This study corroborates the enhanced proficiency of tEEG over conventional EEG in grasping movement decoding, which is crucial for BCI systems. Aligning with prior research \cite{aghaei2020tripolar}, tEEG showed an improved signal-to-noise ratio across various channels. This indicates tEEG's superiority in producing clearer, more reliable EEG signals. Specifically, as it is illustrated in Figure 3, tEEG showed higher median SNRs, especially notable at channel P3, and demonstrated less variability among subjects, suggesting more consistent signal quality. Improved signal-to-noise ratio (SNR) in electroencephalography (EEG) signals can significantly enhance the performance of brain-computer interfaces (BCIs) by increasing the separability of neural patterns associated with different mental tasks or intentions \cite{gu2021eeg}. Higher SNR leads to better feature extraction and classification accuracy, resulting in more reliable and efficient BCI systems. \cite{assi201733} Further, the results in Figure 4  indicate that the tEEG technology exhibits significantly lower functional connectivity between EEG channels compared to conventional EEG, and this finding is in line with prior research \cite{koka2007improvement}. The heatmaps show substantially weaker correlations among channel pairs for tEEG, indicating higher spatial resolution and improved ability to localize and distinguish neural activities originating from different brain regions. This lower functional connectivity and higher spatial resolution afforded by tEEG can potentially enhance the performance of brain-computer interface (BCI) systems by providing more precise and localized information about task-related neural activities. 

In addition, the observed differences in the time-frequency representations between EEG and tEEG, shown in Figure 5, underscore the potential of tEEG to provide significantly improved temporal and spectral resolution. This advancement could facilitate more precise identification of distinct neural patterns associated with various motor tasks, thereby contributing to the improved accuracy of motor task classification across diverse machine learning models employed in this study. Notably, the advantages of tEEG were particularly pronounced in binary classification tasks, where Random Forest and XGBoost classifiers exhibited substantial gains in accuracy when leveraging tEEG signals compared to conventional EEG. Although the Support Vector Machine (SVM) classifier demonstrated robust performance with both EEG modalities, the enhancements afforded by tEEG were less pronounced, suggesting a need for further investigation into optimal feature selection strategies tailored for SVM classifiers when utilizing tEEG data. Importantly, our work achieves up to 91.43\% accuracy for no-movement vs. power grasp, 90.71\% for no-movement vs. precision grasp, 87.88\% for power vs. precision grasp, and 75.97\% for multiclass tasks demonstrating comparable or superior performance to existing research \cite{iturrate2018human, schwarz2017decoding, xu2021decoding}, even with cost-effective EEG headset. The tEEG's advantage is most evident in multiclass classification, illustrated in Figure 6, where it significantly outperforms conventional EEG. Specifically, improvements of 25.45\% for RF, 23.87\% for XGB, and 22.62\% for LDA models highlight tEEG's capability in handling complex neural pattern distinctions crucial for advanced BCI applications, supporting the necessity for high-fidelity signal capture \cite{saha2021progress, roman2012eeg}. These results underscore the robustness of the features extracted using wavelet-based techniques and tEEG signals, which consistently enable superior performance across various machine learning algorithms. This demonstrates the effectiveness of combining tEEG technology with advanced signal processing techniques to enhance the decoding of complex neural activities in grasp-related tasks.   

The findings from this study underscore the significant potential of tEEG technology to improve BCI systems. By offering enhanced signal quality and reliability, higher spatial and temporal resolutions, and improved discrimination of neural activities, tEEG enables more accurate and nuanced control within BCI applications. Furthermore, the improved performance in motor task classification, as demonstrated across various machine learning models, suggests that tEEG could facilitate the development of BCIs with broader applications, from medical rehabilitation to control of external devices. Essentially, these findings highlight tEEG's role in pushing the boundaries of what's achievable with current BCI technology, offering a pathway to more sophisticated and accessible interfaces that could significantly enhance user interaction with technology. 

\section{Limitations and Future Works}
While our findings are promising, it is important to highlight several considerations. The better signal‑to‑noise ratio and sharper spatial resolution of tripolar EEG (tEEG) were already shown in other studies \cite{besio2006tri}, so their presence here is not surprising. We further contributed by applying wavelet-based time-frequency analysis to show how tEEG enhances the spectral and temporal characterization of grasp-related neural dynamics. Additionally, we performed the first direct test of how much these hardware gains translate into higher decoding accuracy for grasp-related brain-computer-interface (BCI) tasks. 

However, several limitations should be acknowledged. First, our study involved only ten healthy adults between 22 and 35 years old, a relatively small sample that does not capture the broader variability found in clinical or aging populations. Second, data were collected from only four scalp locations at a sampling rate of 250 Hz, limiting our ability to detect high-gamma activities and potentially missing important signals from other motor or parietal regions. Third, the task paradigm included only three categories—power grasp, precision grasp, and rest, whereas real-world grasping behaviors involve a wider range of forces, intentions, and movement complexities. 

To address these limitations, future work will pursue several directions. First, we will conduct a larger-scale study that includes a more diverse participant pool spanning different age groups and clinical populations. Second, we aim to expand the tEEG montage beyond four electrodes and increase the sampling rate to capture higher-frequency components, such as high-gamma activity, and to better characterize signals from additional motor and parietal regions involved in grasp planning and execution. Third, we plan to extend the task design to include a wider range of grasp types and force levels (e.g., soft, medium, hard), as well as intentional “null” or “no-go” grasps. These additions will enable the development of a more fine-grained tEEG-based decoding pipeline, better aligned with the complexity of real-world grasping behaviors.

\section{Conclusion}

This work delivers systematic evidence that the well‑known hardware advantages of tripolar EEG translate into meaningful performance gains for practical brain‑computer interfaces. Compared with conventional disc electrodes, tEEG raised signal‑to‑noise ratios, lowered spurious channel‑to‑channel correlations, and, most importantly, boosted classification accuracy by 12–25 percentage points across binary and multiclass grasp‑decoding tasks. These improvements were achieved with just four scalp sites and lightweight machine‑learning models, suggesting that even modest tEEG headsets can outperform far denser conventional arrays.

Beyond technical merit, this study provides a practical roadmap for scaling tEEG from laboratory prototypes to everyday devices. By quantifying SNR gains, spatial‑resolution benefits, and decoder accuracy in a single framework, we furnish design targets for hardware engineers and algorithm developers alike. As future work extends tEEG testing to larger and clinical cohorts, incorporates richer grasp taxonomies, and validates real‑time performance, tripolar electrodes are poised to shift non‑invasive BCIs from proof‑of‑concept toward routine, user‑friendly solutions that bridge the gap between neural intention and the physical world.

% \section{Data availability statement}

% The data can be obtained from the corresponding author upon a reasonable request.

\section{Acknowledgments}
The data can be obtained from the corresponding author upon a reasonable request. Preliminary results of this study were presented at the Society for Neuroscience 2023 Meeting \cite{rabiee2023decoding} (November 2023, Washington DC, USA).

% \section{Conflict of interest}

% The authors declare that there are no relevant conflicts of interest associated with this paper.
% ==============
% # REFERENCES #
% ==============
\printbibliography
% \bibliographystyle{IEEEtran}
% \bibliography{main}

\end{document}